\documentclass[prl,twocolumn,showpacs,amsmath,amssymb,superscriptaddress]{revtex4}
\usepackage{graphicx}
\begin{document}
\title{Exact Landau Levels in Two-Dimensional Electron Systems with Rashba and Dresselhaus Spin-Orbit
Interactions in a Perpendicular Magnetic Field}
\author{Degang Zhang}
\affiliation{Texas Center for Superconductivity and Department 
of Physics, University of Houston, Houston, TX 77204, USA}
\begin{abstract}

We study a two-dimensional electron system in the presence of both Rashba and Dresselhaus spin-orbit interactions in a perpendicular magnetic field. Defining two suitable boson operators and using the unitary transformations we are able to obtain the exact Landau levels in the range of all the parameters. When the strengths of the Rashba and Dresselhaus spin-orbit interactions are equal, a new analytical solution for the vanishing Zeeman energy is found, where the orbital and spin wave functions of electron are separated. It is also shown that in this case the Zeeman and spin-orbit splittings are independent of the Landau level index $n$. Due to the Zeeman energy, new crossing between the eigenstates $|n, k, s=1, \sigma>$ and $|n+1, k, s^\prime =-1, \sigma^\prime>$ is produced at certain magnetic field for larger Rashba spin-orbit coupling. This degeneracy leads to a resonant spin Hall conductance if it happens at the Fermi level. 
 
\end{abstract}
\maketitle

Recently spintronics, operating the spin of charge carriers in semiconductors, has attracted much attention in 
both theory and experiment due to its potential application for information technology [1-7]. One crucial step in spintronics is to produce and efficiently control the spin current. The Rashba spin-orbit interaction [8], which can be adjusted up to 50\% by the application of gate voltages, is used to arrive this goal [9]. The Rashba coupling stems from the lack of structure inversion symmetry, and can be strong in semiconductors (InGaAs, AlGaAs) [10, 11]. Besides the Rashba coupling there also exists a Dresselhaus spin-orbit interaction in semiconductors like InSb/InAlSb [12], which originates from the lack of bulk inversion symmetry. 

The coexistence of the Rashba and Dresselhaus spin-orbit interactions gives rise to some novel spin-dependent phenomena in semiconductors. Recently the spin Hall effect in two-dimensional electron system in the absence of a magnetic field has been studied [13,14]. When the Rashba and Dresselhaus spin-orbit interactions have the same strength, the spin state of the wave functions is independent of the wave vector. The phenomenon was proposed to design a nonballistic Spin-Field-Effect Transistor [15]. The experimental observations of a spin-galvanic effect and spin-orbit coupling weak localization effects have been used to understand the interplay between different spin-orbit types and to obtain the ratio of the Rashba and Dresselhaus spin-orbit coefficients [16,17].

In the presence of a perpendicular magnetic field, the Rashba and Dresselhaus spin-orbit interactions couple all states in each Landau level, and an analytical solution to the full problem has been attempted, but without success [18-23]. As a result, the spin Hall conductance in this system does not get understood. In Refs. [20,21], Shen {\it et al.} investigated transport properties of the electron system with a pure Rashba or Dresselhaus coupling. A resonant spin Hall conductance is found for the Rashba type due to the level crossing induced by the Zeeman energy at the Fermi level. It is very interesting to know whether the resonant spin Hall conductance still exists in the presence of both spin-orbit couplings. When the strengths of both spin-orbit interactions are equal, an exact solution is obtained in the absence of the Zeeman energy [22-25]. However, we shall see below that the energy spectrum is incomplete. The approximate energy spectrum is obtained in Refs. [21-23] by using the perturbation theory. In the present work, we solve analytically this open problem and construct the eigenvalues and eigenstates for such a system so that the spin Hall effect could be investigated in the future. 

The Hamiltonian for a single electron with spin-$\frac{1}{2}$ in a plane under a
perpendicular magnetic field is given by
$$ H=\frac{1}{2m^*}{\bf \Pi}^2-\frac{1}{2}g_s\mu_BB\sigma_z+
V_{SO}({\bf A}),\eqno{(1)}$$
where ${\bf \Pi}={\bf p}+\frac{e}{c}{\bf A}$, $\sigma_i(i=x, y, z)$ are the Pauli matrices 
for electron spin, $g_s$ is the Lande g-factor, 
$\mu_B$ is the Bohr magneton. Here we choose the Landau guage ${\bf A}=yB\hat{{\bf x}}$. 
The spin-orbit interaction has the form
$$V_{SO}({\bf A})=\frac{\alpha}{\hbar}(\sigma_y\Pi_x-\sigma_x\Pi_y)+
    \frac{\beta}{\hbar}(\sigma_x\Pi_x-\sigma_y\Pi_y),\eqno{(2)}$$
where $\alpha$ and $\beta$ represent the Rashba and the Dresselhaus spin-orbit couplings,
respectively. Note that $p_x=k$ is a good quantum number due to $[p_x, H]=0$. 

In order to diagonalize the Hamiltonian (1), we introduce the bosonic annihilation
operator $b_{k\sigma}=\frac{1}{\sqrt{2}l_c}[y+\frac{c}{eB}(k+ip_y)+\sqrt{|a_R a_D|}u_\sigma]$ and the corresponding creation operator $b_{k\sigma}^\dagger=(b_{k\sigma})^\dagger$, with the cyclotron radius $l_c=\sqrt{\frac{\hbar c}{eB}}$, $a_R=\frac{\alpha ml_c}{\hbar^2}$, $a_D=\frac{\beta ml_c}{\hbar^2}$, $u_\sigma=\sigma[1-i sgn(a_R a_D)]$, and the orbital index $\sigma=\pm 1$. 
Here we point out that the constant $\sqrt{|a_R a_D|}u_\sigma$ in the operator $b_{k\sigma}$ plays an important role in solving the Hamiltonian (1). Different the pure Rashba or Dresselhaus coupling case, the orbital space of electron is divided into two infinitely dimensional subspaces described by the occupied number representations $\Gamma_\sigma$ associated with $b_{k\sigma}$ and $b^\dagger_{k\sigma}$ due to this constant term (also see Ref. [23]). When $a_D=a_R$ and $g=0$, the total Hamiltonian (1) can be diagonalized in $\Gamma_\sigma$ by a unitary transformation [22,23]. 
It is surprising that in the special case the eigenstates obtained in Ref. [22-25] are nothing but the basis vectors of two subspaces. This means that the Landau levels decouple. However, in these works, another solution where all the Landau levels couple each other is omitted.
In terms of the bosonic operators $b_{k\sigma}$ and $b_{k\sigma}^\dagger $, the Hamiltonian (1) can be rewritten as
$H=H_{-1}\bigoplus H_1$, and
\begin{widetext}
$$\frac{H_\sigma}{\hbar \omega}=b_{k\sigma}^\dagger b_{k\sigma}+(\sqrt{|a_D a_R|}u_\sigma-i\sqrt{2}a_R\sigma_++\sqrt{2}a_D\sigma_-)b_{k\sigma}^\dagger
+(\sqrt{|a_D a_R|}u_\sigma^*+i\sqrt{2}a_R\sigma_-+\sqrt{2}a_D\sigma_+)b_{k\sigma}
-\frac{1}{2}g\sigma_z+\lambda,\eqno{(3)}$$      
%\end{widetext}
where the cyclotron frequency $\omega=\frac{eB}{mc}$, $\sigma_{\pm}=\frac{1}{2}(\sigma_x \pm i\sigma_y)$,
$\lambda=2|a_Da_R|+\frac{1}{2}$, and $g=\frac{g_sm^*}{2m_e}$
with $m_e$ the free electron mass.   

We can see from Eq. (3) that a state $|m,\uparrow>$ in the basis of $\Gamma_\sigma$ is coupled to $|m+1,\downarrow>$ due to the Dresselhaus coupling, which is further coupled to $|m+2,\uparrow>$ due to the Rashba coupling. Therefore a Landau level is coupled to an infinite number of other Landau levels. This means that
in general the eigenstates for the Hamiltonian (3) in $\Gamma_\sigma$ are convergent infinite series in terms of the parameters $a_D$, $a_R$ and $g$. After a tedious but straightforward calculation, we find that the two-component eigenstate for the nth Landau level with quantum numbers $s$ and $\sigma$ has the form
%\begin{widetext}
$$|n,k,s,\sigma>=\frac{1}{{\cal A}_{ns\sigma}}\sum_{m=0}^{+\infty}
\left(
\begin{array}{cc}
\sqrt{|a_R|}&\sqrt{|a_D|}\Delta_{ns}T^*_\sigma\\
-\sqrt{|a_D|}\Delta_{ns}T_\sigma&\sqrt{|a_R|}
\end{array}\right)
\left (
\begin{array}{c}
\alpha_{ms}^{n\sigma}\\
T_\sigma\beta_{ms}^{n\sigma}
\end{array}\right )\phi_{mk\sigma},$$
$$T_\sigma=\frac{\sqrt{2}}{2}\sigma(sgn a_D+isgn a_R),~~
|{\cal A}_{ns\sigma}|^2=(|a_R|+|a_D|\Delta^2_{ns})\sum_{m=0}^\infty(|\alpha_{ms}^{n\sigma}|^2
+|\beta_{ms}^{n\sigma}|^2),\eqno{(4)}$$ 
%\end{widetext}
where the spin index $s=\pm 1$, $\Delta_{ns}$ is a parameter to be determined below by requiring $\alpha_{ms}^{n\sigma}$ and $\beta_{ms}^{n\sigma}$ to be nonzero, $\phi_{mk\sigma}$ is the eigenstate of the mth Landau level in $\Gamma_\sigma$, i.e. $b_{k\sigma}^\dagger \phi_{mk\sigma}=\sqrt{m+1}\phi_{m+1k\sigma}$, $b_{k\sigma} \phi_{mk\sigma}=\sqrt{m}\phi_{m-1k\sigma}$ and $<\phi_{m^\prime k\sigma^\prime}|\phi_{mk\sigma}>=\delta_{mm^\prime}\delta_{\sigma\sigma^\prime}$. When $m \rightarrow + \infty,  \alpha_{ms}^{n\sigma}=\beta_{ms}^{n\sigma}=0$.

(i) When $|a_D|=|a_R|=a$, the eigenvalue for the $n$th Landau level with $s$ and $\sigma$ is given by
$$E_{ns}\equiv\hbar\omega\epsilon_{ns}=\hbar\omega(n+\frac{1}{2}+2a^2+\frac{1}{2}s\sqrt{g^2+64a^4}).\eqno{(5)}$$
We note that $E_{ns=-1}$ for $g=0$ is nothing but the exact eigenvalue obtained previously [22-25]. Therefore, the exact energy spectrum is incomplete due to the lack of the eigenvalue for $s=1$ in Eq. (5). Obviously, the energy splitting induced by the Zeeman energy and spin-orbit coupling is independent of the Landau level index $n$, i.e. $\Delta E=E_{ns=1}-E_{ns=-1}=\hbar\omega\sqrt{g^2+64a^4}$.

Because the eigenvalues are independent of the parameter $\Delta_{ns}$, for simplicity, we choose $\Delta_{ns}=0$.
For the corresponding eigenstate, we have the recursion relation
%\begin{widetext}
$$[\epsilon_{ns}-m+2a^2-\frac{1}{2}(1-g)]\alpha_{ms}^{n\sigma}=[\epsilon_{ns}-m+2a^2-\frac{1}{2}(1+g)]\beta_{ms}^{n\sigma},$$
$$\sqrt{m}au_\sigma[\alpha_{m-1s}^{n\sigma}+\beta_{m-1s}^{n\sigma}]-[\epsilon_{ns}-m-2a^2-\frac{1}{2}(1-g)]\alpha_{ms}^{n\sigma}
+4a^2\beta_{ms}^{n\sigma}+\sqrt{m+1}au^*_\sigma[\alpha_{m+1s}^{n\sigma}+\beta_{m+1s}^{n\sigma}]=0.\eqno{(6)}$$
Equations (5) and (6) are derived by demanding the coefficient determinant of $\alpha_{ms}^{n\sigma}$ and $\beta_{ms}^{n\sigma}$ to be zero. The same approach is employed in deriving Eqs. (7)-(11) below. 
When $g=0$, from Eq. (6), we have the solution $\alpha_{ns}^{n\sigma}=-\beta_{ns}^{n\sigma}$ and $\alpha_{ms}^{n\sigma}=\beta_{ms}^{n\sigma}=0
(m\not =n)$ for the $n$th Landau level with spin index $s=-1$ in $\Gamma_\sigma$, which recovers the exact eigenstate [23]. We would like to stress that new solution obtained here describes the electron with spin index $s=1$ in $\Gamma_\sigma$. Note that when $g=0$, there is no spin polarization in each Landau level, i.e. $<n,k,s,\sigma|\frac{1}{2}\sigma_z|n,k,s,\sigma>=0$. So it is not expected that there exists an out of plane spin current. In this case, the orbital and spin wave functions of electron are separated, and the spin part $\frac{\sqrt{2}}{2}(1~sT_\sigma)^T$ is independent of the wave vector $k$, which is similar to that observed in the absence of a magnetic field [15].

(ii) When $|a_D|\not =|a_R|$, the reduced eigenvalue is given by
$$\epsilon_{ns}=n-\frac{|a_Da_R|[2|a_Da_R|(1-\Delta_{ns}^2)-g\Delta_{ns}]}{a_R^2-a_D^2\Delta^2_{ns}}
+\frac{1}{2}s\sqrt{[1+\frac{4|a_Da_R|(a_D^2-a_R^2)\Delta_{ns}-g(a_R^2+a_D^2\Delta^2_{ns})}{a_R^2-a_D^2\Delta^2_{ns}}]^2+8n(|a_D|-|a_R|)^2},\eqno{(7)}$$
where $s=1$ for $n=0$ and the parameter $\Delta_{ns}$ is determined by the following equation
$$\sqrt{|a_Da_R|}\{(|a_D|-|a_R|)[2(|a_D|+|a_R|)(|a_R|-|a_D|\Delta_{ns}^2)-g\Delta_{ns}](\epsilon_{ns}-n-\frac{1}{2}+a_D^2+a_R^2)$$
$$+(1+\Delta_{ns})(|a_R|-|a_D|\Delta_{ns})(a_D^2-a_R^2+\frac{1}{2}g)\}=0.\eqno{(8)}$$
If $|a_D|=0$ or $|a_R|=0$, Eq. (8) is automatically satisfied and the eigenvalue (7) becomes that in the presence of a pure Rashba or Dresselhaus spin-orbit interaction [8, 12]. When $g=2|a_R^2-a_D^2|$, we have the analytical expression for the eigenvalue
$$\epsilon_{ns}=n+2|a_Da_R|
+\frac{1}{2}s\sqrt{[1-2sgn(a_R^2-a_D^2)(|a_D|+|a_R|)^2]^2+8n(|a_D|-|a_R|)^2}.\eqno{(9)}$$
Here, we have used $\Delta_{ns}=sgn(a_R^2-a_D^2)$. Note that when $a_D (a_R)\rightarrow a_R (a_D)$, the energy spectrum described by Eq. (9) is identical to that in Eq. (5) with $a_D=a_R$ and $g=0$.

The eigenstate for the $n$th Landau level with $s$ and $\sigma$ is determined by the recursion relation
$$[B_{ns}(\epsilon_{ns}-n-\lambda+1+\zeta_{ns})+|a_Da_R|A_{ns}\eta_{ns}]\alpha_{n-1s}^{n\sigma}=\sqrt{n}u_\sigma^*B_{ns}(B_{ns}-C_{ns})\beta_{ns}^{n\sigma},$$

$$[C_{ns}(\epsilon_{ns}-n-\lambda+\zeta_{ns})+|a_Da_R|A_{ns}\eta_{ns}]\alpha_{ns}^{n\sigma}=\sqrt{n}u_\sigma C_{ns} (C_{ns}-B_{ns})\beta_{n-1s}^{n\sigma},$$

$$\left (
\begin{array}{cc}
\sqrt{m}\sqrt{|a_D a_R|}u_\sigma A_{ns}(C_{ns}-B_{ns})&\sqrt{m}u_\sigma C_{ns}(C_{ns}-B_{ns})\\
-B_{ns}(\epsilon_{ns}-m-\lambda+1+\zeta_{ns})-|a_Da_R|A_{ns}\eta_{ns}&
\sqrt{|a_Da_R|}[A_{ns}(\epsilon_{ns}-m-\lambda+1-\zeta_{ns})+B_{ns}\eta_{ns}]
\end{array}
\right )
\left (
\begin{array}{c}
\alpha_{m-1s}^{n\sigma}\\
\beta_{m-1s}^{n\sigma}
\end{array}
\right )$$
$$=
\left (
\begin{array}{cc}
C_{ns}(\epsilon_{ns}-m-\lambda+\zeta_{ns})+|a_Da_R|A_{ns}\eta_{ns}&
-\sqrt{|a_Da_R|}[A_{ns}(\epsilon_{ns}-m-\lambda-\zeta_{ns})+C_{ns}\eta_{ns}]\\
\sqrt{m}\sqrt{|a_D a_R|}u_\sigma^* A_{ns}(C_{ns}-B_{ns})&\sqrt{m}u_\sigma^* B_{ns}(C_{ns}-B_{ns})
\end{array}
\right )
\left (
\begin{array}{c}
\alpha_{ms}^{n\sigma}\\
\beta_{ms}^{n\sigma}
\end{array}
\right ),\eqno{(10)}
$$
\end{widetext}
where
$$A_{ns}=\frac{(1-\Delta_{ns})(|a_R|-|a_D|\Delta_{ns})}{|a_R|+|a_D|\Delta_{ns}^2},$$
$$B_{ns}=\frac{|a_Da_R|(1-\Delta_{ns}^2)}{|a_R|+|a_D|\Delta_{ns}^2},$$
$$C_{ns}=\frac{a_R^2-a_D^2\Delta_{ns}^2}{|a_R|+|a_D|\Delta_{ns}^2},$$
$$\zeta_{ns}=\frac{\frac{1}{2}g(|a_R|-|a_D|\Delta_{ns}^2)+4|a_Da_R|(|a_D|+|a_R|)\Delta_{ns}}{|a_R|+|a_D|\Delta_{ns}^2},$$
$$\eta_{ns}=\frac{2(|a_R|+|a_D|)(|a_R|-|a_D|\Delta_{ns}^2)-g\Delta_{ns}}{|a_R|+|a_D|\Delta_{ns}^2}.
\eqno{(11)}$$
We note that when $a_D=0$ or $a_R=0$, the corresponding eigenstates presented in Refs. [8,12,20-25] can be easily obtained from Eq. (10).

Fig. 1 shows several low-lying Landau levels as functions of $|a_R|$ for $|\beta|=0.2|\alpha|$, $0.6|\alpha|$, and $1.2|\alpha|$ and $g=0$ and 0.1. When $|a_R|$ is small, the energy levels $E_{ns=1}$ and $E_{n+1s=-1}$ are approximately degenerate for $g=0$ [see Fig. 1(a), (c) and (e)]. If $g\not =0$, the approximate degeneracy is lefted in Fig. 1(b), (d), and (f). However, the Landau levels shift in an opposite way for $|a_R|>|a_D|$ and $|a_R|<|a_D|$. When $|a_R|>|a_D|$ new crossing between $|n,k,s=1,\sigma>$ and $|n+1,k,s^\prime =-1,\sigma^\prime>$ pointed by arrows is produced in Fig. 1(b) and (d) while there is no new mixing for $|a_R|<|a_D|$ in Fig. 1(f). This new degeneracy induced by the Zeeman energy leads a resonant spin Hall conductance if it happens at the Fermi level, similar to the case in the presence of the pure Rashba spin-orbit interaction [20,21]. It is expected that the resonant spin Hall conductance can be applied in device design. The degenerate point at the Fermi level is easily controlled by the gate voltage, which changes the Rashba coupling strength. We note that with increasing the Dresselhaus coupling, the resonance is moved to the lower magnetic fields.

In summary, we have analytically derived the complete energy spectrum in 2D electron systems in the presence of both Rashba and Dresselhaus spin-orbit interactions under a perpendicular magnetic field. The eigenvalue for the $n$th Landau level with $s$ and $\sigma$ is determined by the parameter $\Delta_{ns}$, which satisfies the highly nonlinear equation (8) for $|a_D|\not =|a_R|$. For $|a_D|=|a_R|$, we obtain a new analytical solution when the Zeeman energy vanishes. Each Landau level possesses two-fold degeneracy because it is independent of the orbital index $\sigma$. We also note that the eigenvalues depend only on the absolute values of the Rashba and Dresselhaus spin-orbit coefficients rather than their signs. The corresponding eigenstates are convergent infinite series except those at $|a_R|=|a_D|$, $g=0$ and $s=-1$. Once the energy spectrum is known, we can use it to calculate physical properties of the system, such as charge and spin Hall conductance. The work in that direction is in progress.  

\begin{figure}
\rotatebox[origin=c]{0}{\includegraphics[angle=0, 
           height=1.49in]{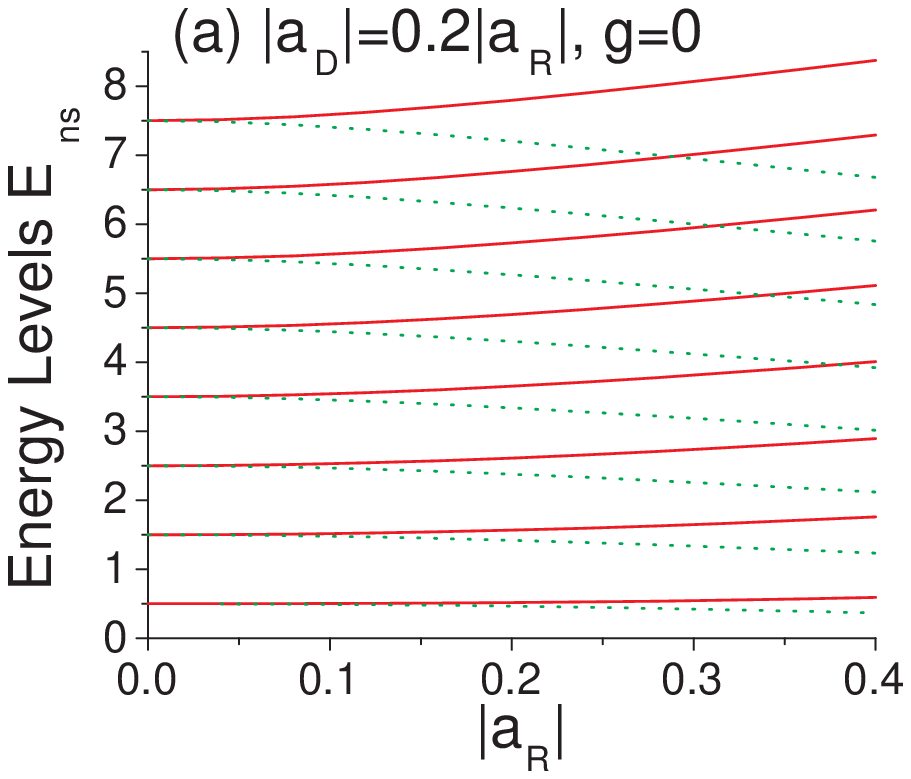}}
\rotatebox[origin=c]{0}{\includegraphics[angle=0, 
           height=1.49in]{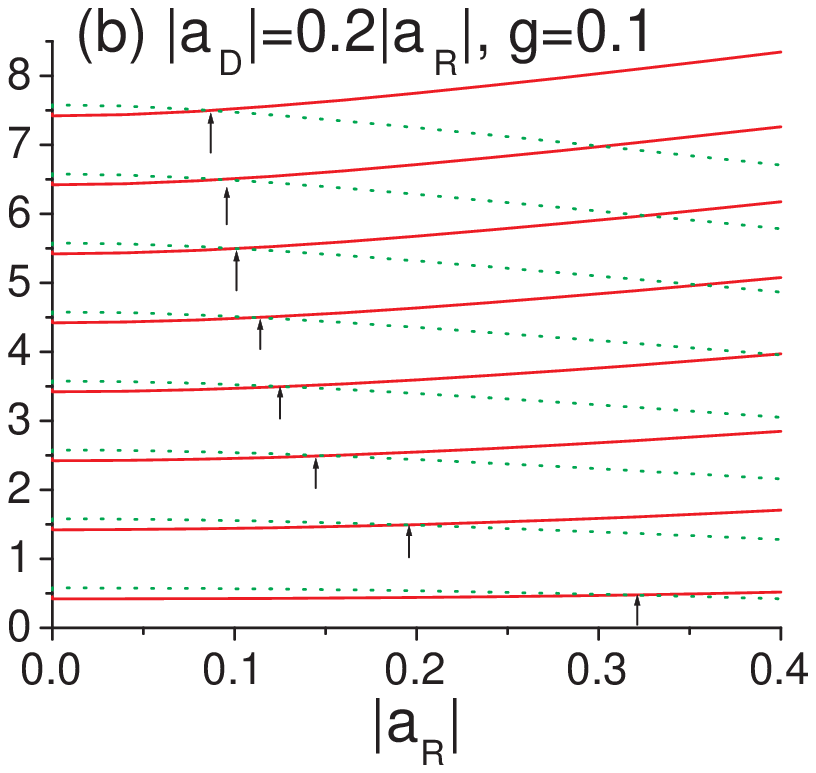}}
\rotatebox[origin=c]{0}{\includegraphics[angle=0, 
           height=1.49in]{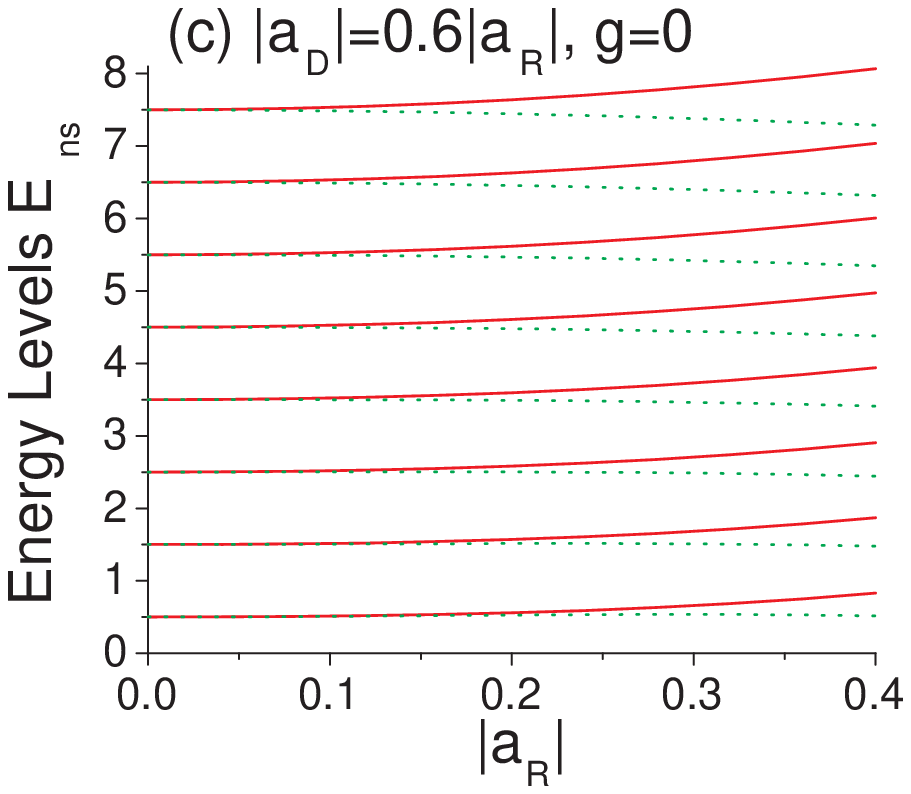}}
\rotatebox[origin=c]{0}{\includegraphics[angle=0, 
           height=1.49in]{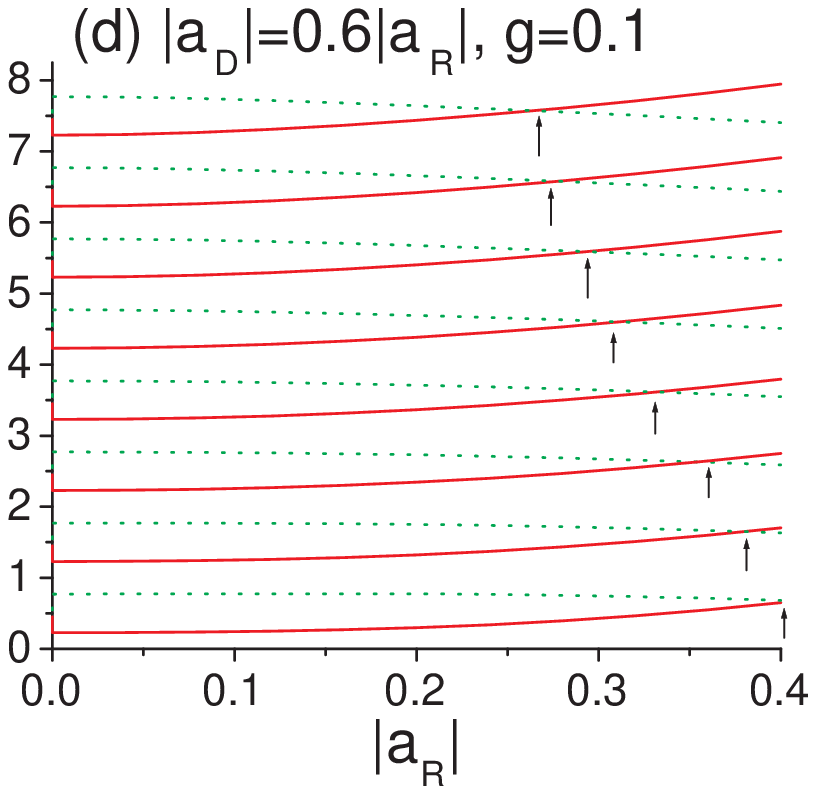}}
\rotatebox[origin=c]{0}{\includegraphics[angle=0, 
           height=1.49in]{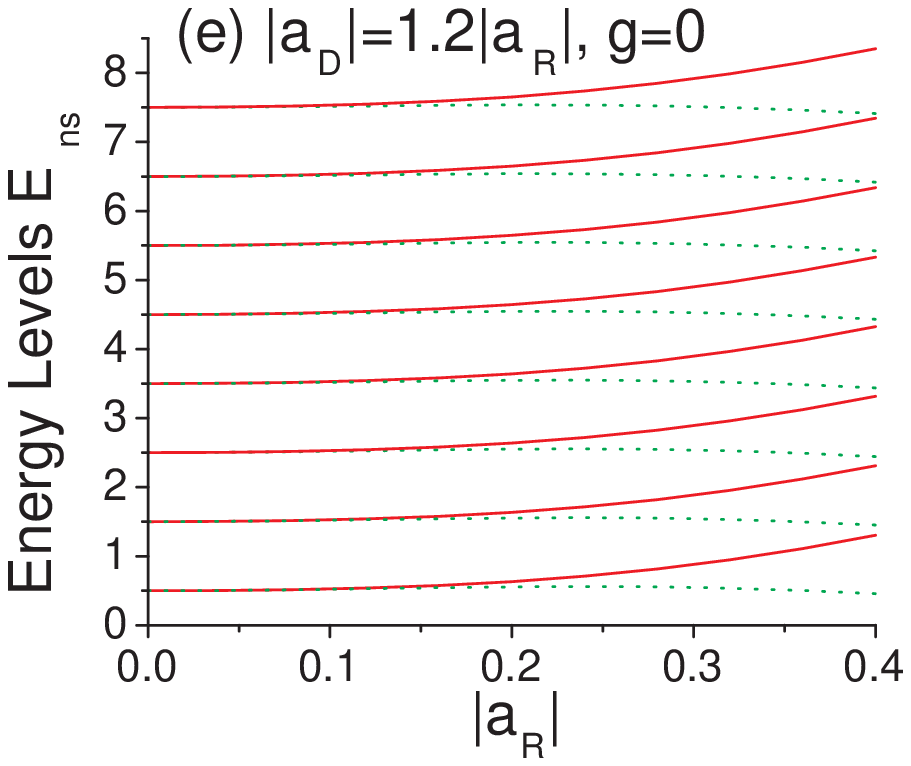}}
\rotatebox[origin=c]{0}{\includegraphics[angle=0, 
           height=1.49in]{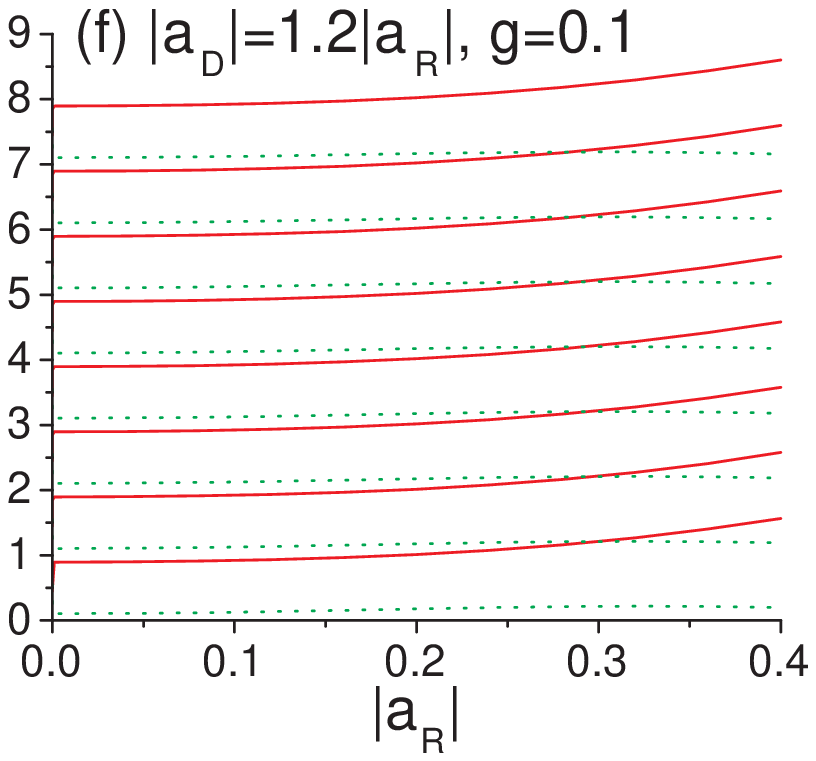}}           
\caption {(Color online) Landau levels in unit of $\hbar\omega$ as a function of the dimensionless parameter $|a_R|$. The parameters $|a_D|$ and $g$ are shown in the panels. Solid, $s=1$; dot, $s=-1$.}
\end{figure}

The author would like to thank Prof. C. S. Ting and Prof. C.-R. Hu for useful discussions. This work was supported by the Texas Center for Superconductivity at the University of Houston 
and by the Robert A. Welch Foundation under grant No. E-1146.

%\end{multicols}

\end{document}